\documentclass[twocolumn]{webofc}

\usepackage[varg]{txfonts}   
\usepackage{graphicx}
\graphicspath{{./images/}}

\begin{document}
\title{Speak with signs: Active learning platform for Greek Sign Language, English Sign Language, and their translation}

\author{\firstname{Maria} \lastname{Papatsimouli}\inst{1} \and
        \firstname{Lazaros} \lastname{Lazaridis}\inst{1}   \and
        \firstname{Konstantinos-Filippos} \lastname{Kollias}\inst{1} \and
        \firstname{Ioannis} \lastname{Skordas}\inst{1} \and
        \firstname{George F.} \lastname{Fragulis}\inst{1}
        \fnsep\thanks{\email{gfragulis@uowm.gr}}
}

\institute{Laboratory of Robotics, Embedded and Integrated Systems, \\ 
	Dept. of Electrical and Computer Engineering, \\
	University of Western Macedonia, Hellas}

\abstract{%
  Sign Language is used to facilitate the communication between Deaf and non-Deaf people. It uses signs-words with basic structural elements such as handshape, parts of face, body or space, and the orientation of the fingers-palm. Sign Languages vary from people to people and from country to country and evolve as spoken languages. In the current study, an application which aims at Greek Sign Language and  English Sign Language learning by hard of hearing people and talking people, has been developed. The application includes grouped signs in alphabetical order. The user can find Greek Sign Language signs, English sign language signs and translate from Greek sign language to English sign language. The written word of each sign, and the corresponding meaning are displayed. In addition, the sound is activated in order to enable users with partial hearing loss to hear the pronunciation of each word. The user is also provided with various tasks in order to enable an interaction of the knowledge acquired by the user. This interaction is offered mainly by multiple-choice tasks, incorporating text or video. The current application is not a simple sign language dictionary as it provides the interactive participation of users. It is a platform for Greek and English sign language active learning.
}

\maketitle

\section{Introduction}
\label{intro}
Language is an innate mechanism that humans develop \cite{Malmberg2012}. People with hearing problems also seek a way to communicate and need to develop a language that is directly accessible and effective for them. One such language is sign language \cite{Kushwah2017}. Sign languages are the only languages that Deaf people can use in order to communicate in a natural, effortless, easy, reciprocal and effective way. In Greece, Greek Deaf people use the Greek Sign Language (GSL), which is their natural language, as it is used not only by the majority of them but also by their hearing children, as well as by professionals and experts who work with deaf people. 
Nowadays, the use of Information and Communication Technologies in everyday life has shown an increasing trend and has helped many people in their everyday life. The deaf/hard of hearing people could not have been unaffected by these rapid changes. The use of technology has the effect of reducing isolation, increasing independence, and offering educational, economic, and social opportunities to  deaf/hard of hearing people \cite{MaioranaBasas2014}.

\subsection{Deaf and Sign Languages}
According to \cite{Woodward1972}, Deaf with ‘D’ are those deaf/hard of hearing people who belong to the Deaf community and use sign language in order to communicate, while deaf with ‘d’ are those who are hard of hearing and do not necessarily need sign language as a communication tool. Sign language is the natural language of Deaf people and not just an artificial communication system. Each country has its own Sign Language with structural features that differ from spoken languages. The gestures consist of regular structures and semantics that correspond to spoken languages \cite{Stokoe1980}. Deaf people of each country use their own Sign Language \cite{Panagiotakopoulos2003}.

\subsection{Greek Sign Language}
Greek Sign Language is the mother tongue of Greek Deaf people. Sign Language had been sidelined in many European countries, for many years. Greek Sign Language is a complete and independent language, recognized as "a non-written language with all the linguistic phenomena observed in the spoken languages” (grammar, syntax, dictionary, phonology). In addition, the natural language of Deaf people presents elements of morphology, syntax, semantics, and pragmatics, while the linguistic system of phonology is replaced by the corresponding italics \cite{Aarssen2018}.
In sign language, the combination of handshape with other elements, such as direction, position and movement, gives a specific meaning to a word. More specifically, direction has to do with the orientation that the palm takes, position shows the point where the hand is placed in relation to the body and movement shows other syntactic information such as the subject-object agreement \cite{Ackovska2012}, \cite{MaioranaBasas2014}, \cite{Papaspyrou2003}, \cite{Sandler2001}, \cite{Valli2000}. One or both hands are utilized in order to express the sign,  while making the necessary movements. The signs that are rendered in this way are the main elements that distinguish sign language from spoken language \cite{Sapountzaki2015}. 
Finally, there is a Finger Alphabet that is a morphological element of sign language \cite{Aarssen2018}. Finger alphabet represents Greek alphabet of spoken language and differs from signs. A Deaf person can use this alphabet in order to spell some Greek words as they are in a visualized way, or form names with his fingers \cite{Marschark2007}.

\subsection{Education and Information \& Communication Technologies (ICT)}
During the training process, the use of tools and software for educational purposes which utilize multimedia and internet technologies is proposed. In this way, students are enabled to develop and adapt the knowledge acquired at school to the modern educational environment, and have the opportunity to collect, represent, analyze, transfer, and utilize information. Mental processes and knowledge acquisition \cite{Jonassen2000} are utilized through an educational environment which results in the development of new skills and abilities. Therefore, a new learning culture is created  and leads to a meaningful relationship between knowledge and its construction.

\section{Related Work - Applications of Sign language in Greece and Worldwide}

One of the most fundamental features of software applications is interaction. Interaction helps each user to be transformed from a passive recipient to an active member of learning process that keeps his interest undiminished.
Sign Language is a visual language, and with the contribution of video, it can be included in any application in order to transfer information and provide  deaf/hard of hearing people with easy access to knowledge \cite{Kakoty2018},\cite{Kim2019}.

The majority of applications that have been developed to date are related to learning sign language and  translating from signs to text or spoken language.
For example, the following are some applications from Greece and Worldwide:

\begin{enumerate}
	\item Greek Sign Language: The web application has been operating with free access since 2016 and it was developed by the University of Patras. In this application, users can find signs of the basic vocabulary for everyday use. The application is aimed at children and adults who want to learn Greek sign language. However, there is no interaction between the user and the content \cite{Various2020a}.
	\item Greek Sign Language Center: It is a free access web application. It has been developed by the Greek Sign Language Center and contains alphabetically ordered videos for sign learning. The platform is addressed to children and adults and provides quizzes for practice that contain videos with multiple-choice questions. Users can see a video and then choose the right answer.
	\item DIOLKOS Software: Educational software for training in computers operation with terminology in Greek Sign Language, Greek, and English. Developed in 2006.
	\item LEARNING MEANINGS Software: It is a teaching environment for Greek Sign Language (GSL) vocabulary developed in 2013. This software had been addressed to students in the first grades of Primary School. The arranging of its contents based on the principles, characteristics and rules regulate the vocabulary of the language.
	\item CHILDREN'S DICTIONARY OF GREEK SIGNIFICANT LANGUAGE Software: It included videos with Greek signs translated into the corresponding Greek words. It had been addressed to kindergarten and first grades of primary school children. Developed in 2001.
	\item Greek Sign Language Courses: This application contains words translated into GSL. It includes basic signs, complex signs, synonyms and antonyms, the finger alphabet and vocabulary groups. It is available for free and is addressed to all age groups \cite{Various2020}.
	\item ASL-LEX: It is an online application that displays signs of American Sign Language. Users can see the frequency of  use,  lexical properties, the phonological coding, and other characteristics of each sign. Also, they have the aability to search for the written word of each sign to display \cite{Various2020d}.
	\item preadTheSign:	It is an online application that gives signs in many different sign languages. For example, English of the United States, English of India, German, Greek, Japanese, etc. This application groups its content in terms of subject and not alphabetically. Users can interact with the content by using 360 degrees images where, there are points of interest that the user can click and see the corresponding signs \cite{Various2020e}.
	\item Handspeak: It is an online application that displays the signs in English Sign Language but also in American Sign Language. This application gives the content in alphabetical order and is addressed to groups of all age. The synonyms of each word are displayed and the user can see them by clicking on the corresponding word.Also, it is possible to display videos that show stories in sign language(storytelling in sign language) \cite{Lapiak2020} .
\end{enumerate}

\section{Application Description}

The proposed application (http://signlanguage.groupdvs.com/) has been developed aiming at the acquisition of a basic Greek Sign Language vocabulary that is used daily. Particularly, it is addressed to children and adults who want to learn Greek Sign Language(GSL), deaf children who do not have prior knowledge of GSL, parents of deaf children and  hearing children who want to learn GSL. It has to be mentioned that relevant applications with a purely educational character both in Greece and internationally, do not exist. As it was mentioned above, there are some dictionarylike applications, which are used for Sign Language learning. Most of the aforementioned applications do not provide user interaction for practice and in-depth learning of sign language. The application was designed as an autonomous platform for tele-education and was created under the philisophy of open source software. In recent years our research team has developed a number of applications using open-source programming languages and tools such as PHP, MySQL and WordPress \cite{Fragulis2018}, \cite{Lazaridis2016}, \cite{Lazaridis2019}, \cite{Michailidi2020}, \cite{Papatsimouli2020}, \cite{Skordas2017}.

\begin{figure}[h]
	\includegraphics[width=0.5\textwidth]{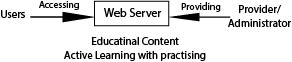}
	\caption{Main Workflow}
	\label{lab-f2}       
\end{figure}

\begin{figure}[h]
	\includegraphics[width=0.5\textwidth]{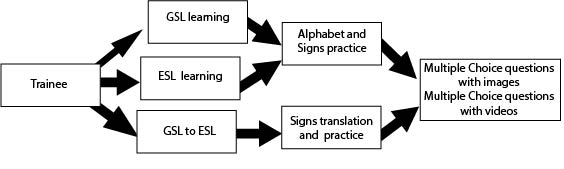}
	\caption{Detailed Application Workflow}
	\label{lab-schema}       
\end{figure}

In the figures (\ref{lab-f2})-(\ref{lab-schema}), we give the main \& detailed Application workflow. The administrator can upload training material to the application and users have access to this material and can practice aiming at acquiring knowledge through their active participation (active learning).

\subsection{Requirements' Analysis}

The main points of the requirements analysis found for online educational applications are the following:
\begin{itemize}
	\item Vocabulary categorization into semantic sections for the facilitation of users
	\item Videos are the best format for use, except for the display of finger alphabets where images are appropriate, too
	\item There should be a connection between each word and each meaning in a visual form (image/video)  in order to support users who have a poor knowledge of Greek written language (e.g. children or adults with low educational level)
	\item The interface should be as simple as possible and easy for users
	\item In this application, the user will be able to see displays of signs/videos. In addition, practice will provide users with increased and appropriate knowledge through active learning
	\item The repetition of videos should be possible
	\item Audio integration should be available in order to support hard of hearing people
	
\end{itemize}

\subsection{User Interface and Functionality}

User’s registration is not required in order to access the application.
During the operation of the application, there is immediate feedback for every action of the user.
\begin{figure}[h]
	\includegraphics[width=0.5\textwidth]{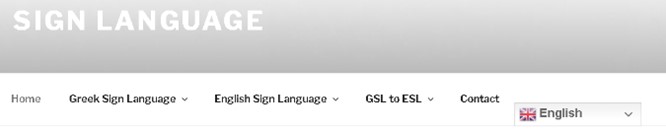}
	\caption{Home page}
	\label{fig-f1}       
\end{figure}
\subsubsection{HomePage}
The Home page (http://signlanguage.groupdvs.com/) consists of 5 different menus: Home, Greek Sign Language, English Sign Language, GSL to ESL, Contact. The content is automatically translated depending on the language chosen by each user (English, Chinese, German, etc.) The Home page provides information about sign language and the Contact page allows the user to contact the administrator.

\begin{figure}[h]
	\includegraphics
	[width=.5\textwidth,
	height=4cm,
	keepaspectratio,]{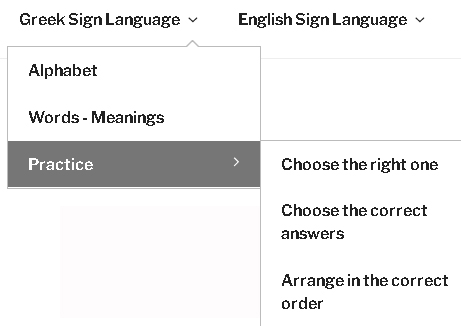}
	\caption{Greek Sign Language submenu}
	\label{fig-2}       
\end{figure}

 \subsubsection{Greek Sign Language Menu}
 In this section, the user can learn the Greek Sign Language alphabet, search for signs which are categorized in alphabetical order and practice on the acquired knowledge. English Sign Language menu has the same structure as the Greek Sign Language one. Specifically, it contains alphabet presentation, search for signs by their first letter, and practice.

 \begin{figure}[h]
 	\includegraphics
 	[width=.5\textwidth,
 	height=2cm,
 	keepaspectratio,]{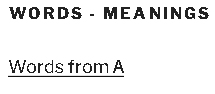}
 	\caption{Example of an alphabetical display of signs per letter}
 	\label{fig-3}       
 \end{figure}
 
\begin{figure}[h]
 	\includegraphics
 	[width=.5\textwidth,
 	height=8cm,
 	keepaspectratio,]{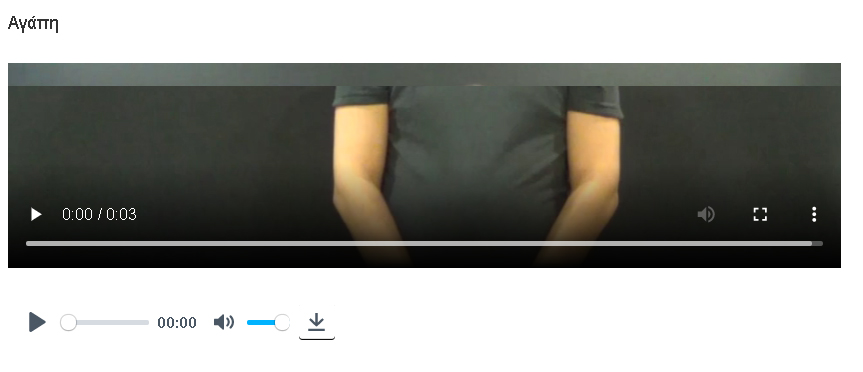}
 	\caption{Example of selected word display}
 	\label{fig-4}       
 \end{figure}

The user selects the category that he wants, and all the words that start with the selected letter are displayed. As it is presented in figure (\ref{fig-4}), for each selected word the written text, the sign (in video format), and the pronunciation are displayed, for supporting the hard of hearing people, who do not have complete hearing loss.

 \subsubsection{Translation of Greek Sign Language to English Sign Language}
In this menu, the user can translate Greek sign language signs into English sign language and exercise in the translation of signs, as well.
 
  \begin{figure}[h]
 	\includegraphics
 	[width=.5\textwidth,
 	height=3cm,
 	keepaspectratio,]{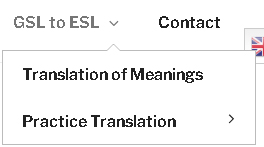}
 	\caption{Greek sign language to English sign language}
 	\label{fig-5}       
 \end{figure}
 
\begin{figure}[h]
 	\includegraphics
 	[width=.5\textwidth,
 	height=3cm,
 	keepaspectratio,]{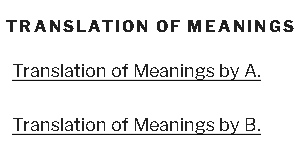}
 	\caption{Group content to view}
 	\label{fig-6}       
 \end{figure}

The first option is translation of Greek sign language signs into English sign language and vice versa. The signs are categorized according to Greek Alphabet, and the user can select the word group to be displayed. At this point, the written text, the sign (in video format), and the pronunciation are displayed both in Greek and English (Figures \ref{fig-5}-\ref{fig-6}).

\subsubsection{Practice content task}
In this task, users can exercise the Greek finger alphabet with Figures. Users see the letters and then enter the correct answer. At the end of the task, users can see the achieved results on the screen.

 \begin{figure}[h]
 	\includegraphics
 	[width=.5\textwidth,
 	height=6cm,
 	keepaspectratio,]{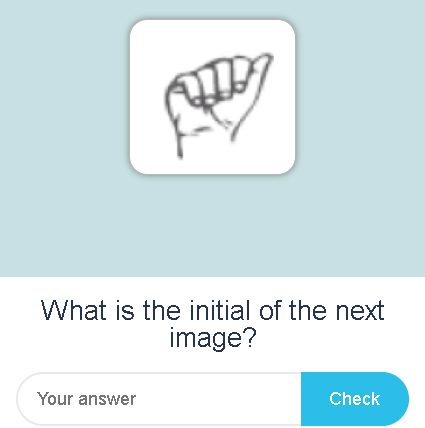}
 	\caption{Choose the right answer}
 	\label{fig-7}       
 \end{figure}
 
\begin{figure}[h]
 	\includegraphics
 	[width=.5\textwidth,
 	height=4cm,
 	keepaspectratio,]{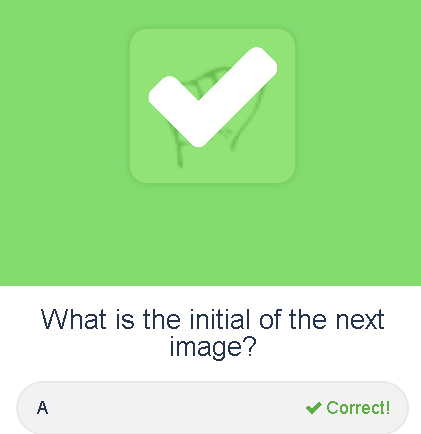}
 	\caption{Correct answer}
 	\label{fig-8}       
 \end{figure}

\begin{figure}[h]
 	\includegraphics
 	[width=.5\textwidth,
 	height=4cm,
 	keepaspectratio,]{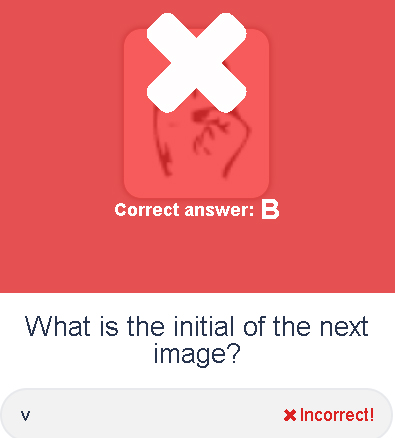}
 	\caption{Wrong answer}
 	\label{fig-9}       
 \end{figure}

The user enters the answer in the box, and then he presses the Check button (Figure \ref{fig-7}) . The answer is corrected automatically depending on the result (Figures \ref{fig-8}-\ref{fig-9}). After completing the task, users can see the overall results (Figure \ref{fig-2})).

 \begin{figure}[h]
 	\includegraphics
 	[width=.5\textwidth,
 	height=8cm,
 	keepaspectratio,]{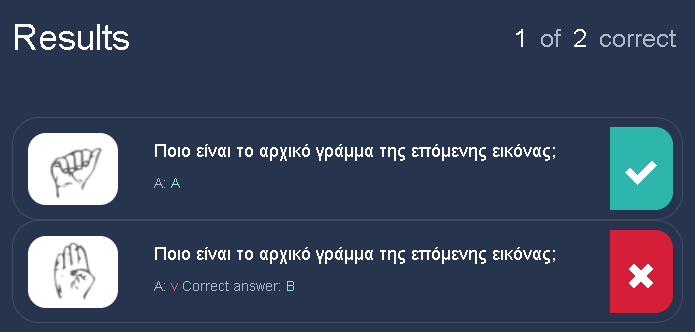}
 	\caption{Final results presentation}
 	\label{fig-2}       
 \end{figure}

\subsubsection{Choose the correct answers for Greek Sign Language practice}
\begin{figure}[h]
	\includegraphics
	[width=.5\textwidth,
	height=5cm,
	keepaspectratio,]{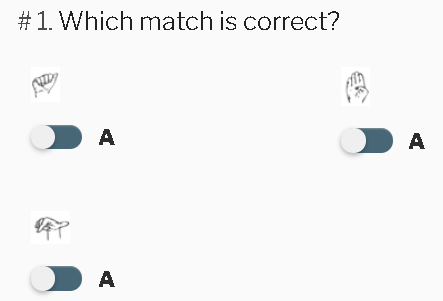}
	\caption{Select the correct match}
	\label{fig-11}       
\end{figure}

This task is about Greek Finger Alphabet practice. Various options are presented and the user chooses the right combination of a letter and a Figure. When the test is finished, the achieved results are displayed (Figure \ref{fig-11}).

\subsubsection{Arrange in correct order task in the Greek finger alphabet}
\begin{figure}[h]
	\includegraphics
	[width=.5\textwidth,
	height=8cm,
	keepaspectratio,]{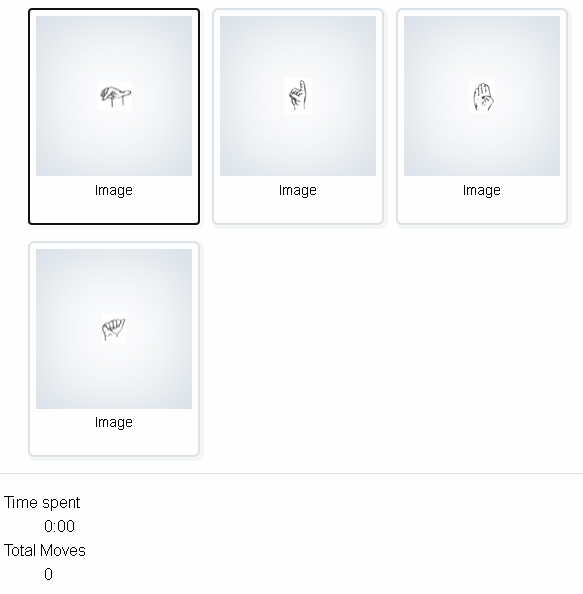}
	\caption{Arrange in correct order}
	\label{fig-12}       
\end{figure}

In this task, the user places the Figures in correct order so that the finger alphabet appears in alphabetical order(Figure \ref{fig-12}).

Also the users can see the time that they spent on doing the task, and the total moves that were made. In addition, they can press either Check button to complete the task or Show Solution to see the solution.

\subsubsection{English finger alphabet recognition task}

\begin{figure}[h]
	\includegraphics
	[width=.5\textwidth,
	height=10cm,
	keepaspectratio,]{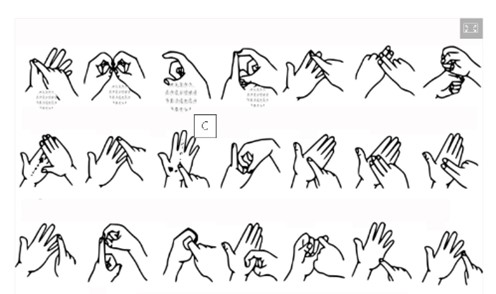}
	\caption{Recognition of English Sign Language letters}
	\label{fig-13}       
\end{figure}

Here, the user recognizes the English finger alphabet (Figure \ref{fig-13}). The user sees  figures depicting the English Sign Language letters and by moving the cursor over them, the English characters corresponding to them will appear.

\subsubsection{Videos with Multiple Choice questions}
\begin{figure}[h]
	\includegraphics[width=0.5\textwidth]{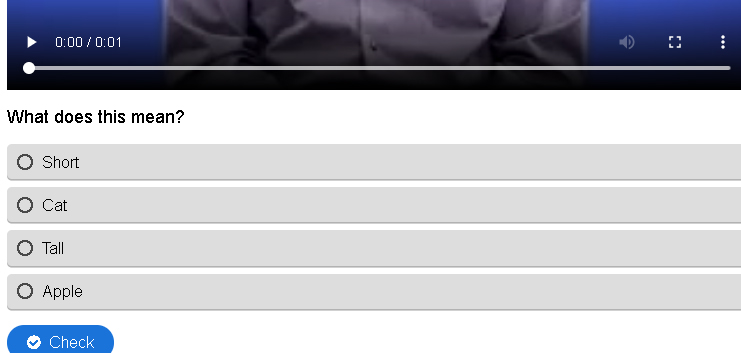}
	\caption{Videos with Multiple Choice questions task}
	\label{fig-14}       
\end{figure}

In this task, the user can see the signs in video format (Figure \ref{fig-14})., choose the correct answer from the available options that are shown and press the Check button to check the answer.

\subsubsection{Choose the first letter of the word}
\begin{figure}[h]
 	\includegraphics
 	[width=.5\textwidth,
 	height=10cm,
 	keepaspectratio,]{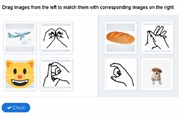}
 	\caption{Choose the first letter of the word task}
 	\label{fig-15}       
 \end{figure}

\begin{figure}[b]
	\includegraphics
	[width=.5\textwidth,
	height=10cm,
	keepaspectratio,]{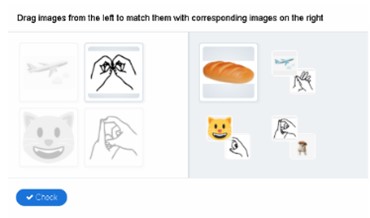}
	\caption{Match the letters with the right Figure}
	\label{fig-16}       
\end{figure}

\begin{figure}[h]
	\includegraphics
	[width=.5\textwidth,
	height=10cm,
	keepaspectratio,]{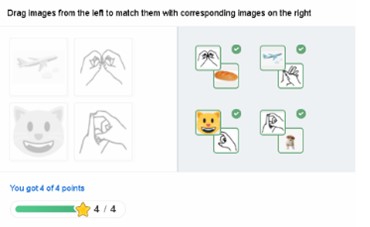}
	\caption{Feedback given}
	\label{fig-17a}       
\end{figure}

In this task, the initial letters in English Sign Language and pictures that start with the corresponding letters are given (Figures \ref{fig-15} - \ref{fig-17a}). The user makes the right combinations, and feedback appears by pressing the Check button.

\subsubsection{Practice in storytelling}
\begin{figure}[b]
	\includegraphics
	[width=.5\textwidth,
	height=10cm,
	keepaspectratio,]{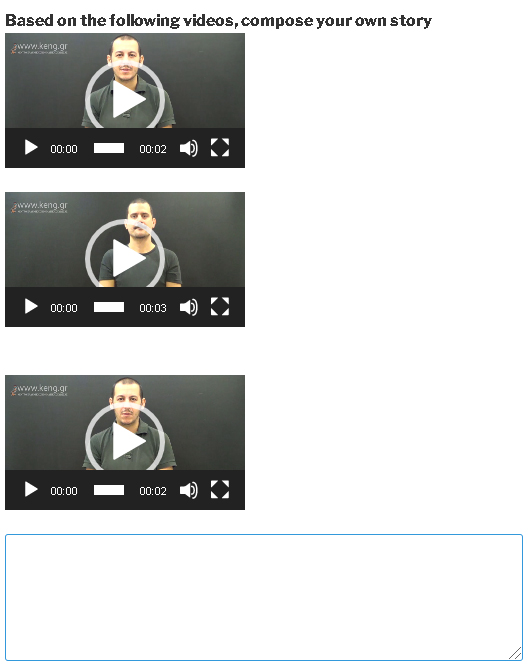}
	\caption{Practice in storytelling}
	\label{fig-17}      
\end{figure}

In this task, the user  watches some videos including signs and then uses these signs in order to write a short story (Figure \ref{fig-17}). It is one of the tasks that contribute to the development of imagination of users as everyone can write his/her own story without being right or wrong. It is a non-linear task that helps the development of written narrative speech.

\subsubsection{Memory cards task}
\begin{figure}[h]
	\includegraphics
	[width=.5\textwidth,
	height=5cm,
	keepaspectratio,]{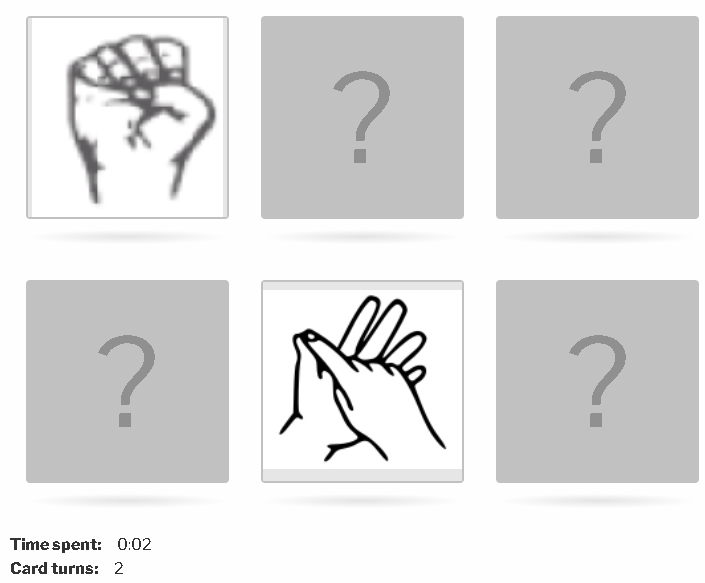}
	\caption{Memory Cards Task}
	\label{fig-18}       
\end{figure}

In this task, the user matches the letters of the Greek Sign Language with the  English Sign Language ones (Figure \ref{fig-18}) . The user can see the moment and the amount of times the cards were turned.

\subsubsection{Interactive Videos Task}
\begin{figure}[h]
	\includegraphics
	[width=.5\textwidth,
	height=10cm,
	keepaspectratio,]{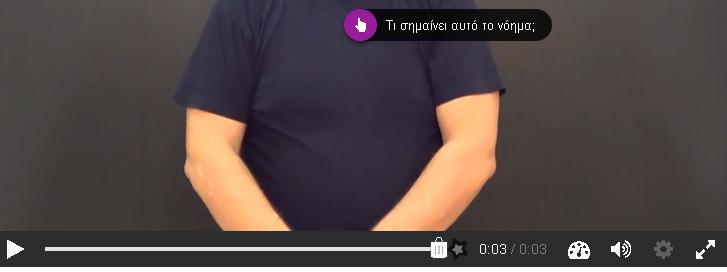}
	\caption{Interactive videos task}
	\label{fig-19}       
\end{figure}

In this task, a sign is shown on a video and then it stops (Figure \ref{fig-19}). The user needs to answer what this sign means by clicking on an active point on the video, and choosing one of the available options that are shown. The user can check his answer by pressing the Check button.

\begin{figure}[b]
	\includegraphics
	[width=.5\textwidth,
	height=10cm,
	keepaspectratio,]{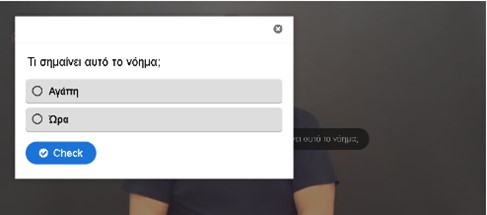}
	\caption{Answer selection in interactive videos task}
	\label{fig-20}       
\end{figure}

 \section{Conclusions}
 Sign languages are similar to spoken languages and it is the communication system that is used in deaf communities. It is acquired during childhood without being instructed, achieves the same social and mental functions as spoken languages and can be interpreted in real-time \cite{Cormier2006}. The introduction of ICT in education can bring important results to the educational process. Moreover, the appropriate introduction of educational methods in online platforms can give the best results in knowledge acquisition and make the educational process more interesting. In the current application, users can learn both Greek and English sign language and translate them, as well. The educational content of all these categories has been grouped in alphabetical order in order to enable the user to easily find it. Each word has been presented in written text both in Greek and English, in sign language in video format and its pronunciation has been presented as well in order to support the hard of hearing people without total hearing loss.  The current platform -using only open source software- is not just a simple dictionary of signs. Users can interact with the educational content and actively participate in the educational process through active learning. The user can exercise in various types of tasks and receive feedback. User’s interaction is based on Figures, depicting the finger alphabet and videos displaying signs.  The most important and innovative tasks that were used for sign language learning are:

 \begin{itemize}
 	\item Arrange in correct order, in which the user places the letters of the finger alphabet in  right order
 	\item Memory Cards where users have to match each Greek finger alphabet letter with the English finger alphabet one
 	\item Interactive videos, which show a sign and the user chooses what this sign means by choosing one of the available options
 	\item Choose the first letter of the word, where initial letters in English sign language and pictures that start with the corresponding letters are given. The user recognizes the letters of the finger alphabet and makes the right combinations.
 	\item Storytelling where videos are displayed, and the user can write his own story based on the signs that he just saw.
 \end{itemize}




%

%

%
%

\end{document}